\documentclass[pre,aps,preprint,showpacs,preprintnumbers,amsmath,amssymb,secnumroman,eqsecnum]{revtex4}

\usepackage{amsmath}
\usepackage{amssymb}
\usepackage{graphicx}
\usepackage{dcolumn}
\usepackage{bm}

\newcommand{\beq}{\begin{equation}}
\newcommand{\eeq}{\end{equation}}
\newcommand{\beqa}{\begin{eqnarray}}
\newcommand{\eeqa}{\end{eqnarray}}
\newcommand{\vc}[1]{\mbox{\boldmath $#1$}}

\newcommand{\vol}[1]{{\bf #1}}

\begin{document}


\title{Derivation of the Love equation for the charge density of a circular plate condenser}

\author{B. U. Felderhof}

 \email{ufelder@physik.rwth-aachen.de}
\affiliation{Institut f\"ur Theoretische Physik A\\ RWTH Aachen University\\
Templergraben 55\\52056 Aachen\\ Germany\\
}%

\date{\today}

\begin{abstract}
The electrostatic problem of two coaxial parallel charged disks of different radii in infinite space is solved by expansion in terms of complex potentials. For perfectly conducting disks a pair of coupled integral equations is derived for the two weight functions of the expansion. The weight function on each disk is linearly related to its radially symmetric charge density. If the radii are equal and the disks are oppositely charged the pair of equations reduce to Love's integral equation for the circular plate condenser.
\end{abstract}

\pacs{02.30.Em, 02.60.Lj, 41.20.Cv}
\maketitle

\section{\label{I}Introduction}

The electrostatic problem of the circular plate condenser has received much attention ever since the early work of Maxwell \cite{1} and Kirchhoff \cite{2}. For the case of two equal coaxial circular conducting disks Nicholson \cite{3} solved Laplace's equation in spheroidal coordinates, and expressed the electrostatic potential as a series of spheroidal harmonics with coefficients satisfying an infinite system of linear equations. His expression for the potential was shown by Love \cite{4} to be meaningless, since it contained a non-convergent integral. Based on Nicholson's work, Love derived an integral equation for a weight function related to the surface charge density on each of the plates. The equation can be solved numerically to find the value of the capacitance for any separation distance of the plates.

Love's derivation of his integral equation was quite complicated. A simpler derivation, based on the solution of Laplace's equation in cylindrical geometry by an expansion in terms of Bessel functions, was provided by Sneddon \cite{5}. His derivation involves Hankel transforms and dual integral equations.

Another derivation of the Love equation based on an expansion of the electrostatic potentials in terms of Bessel functions, closely related to Sneddon's work, was presented by Carlson and Illman \cite{6}. That such an expansion can be hazardous, unless proper care is taken with the formulation of dual integral equations, is evident from the work of Atkinson et al. \cite{7}, whose expression for the potential of the circular plate condenser was shown by Love \cite{8} to be incorrect.

Love himself took the view that it is sufficient to establish that an expansion of the electrostatic potential in terms of complex potentials satisfies all requirements. In Love's presentation the weight function in the expansion does not have an obvious physical meaning. He showed only that its integral is proportional to the total charge. A gap in his derivation was closed by Reich \cite{9}. The expansion in terms of complex potentials was studied independently by Green and Zerna \cite{10}, who showed for a single disk how to find the surface charge density  from the weight function. Their work was discussed by Sneddon \cite{5}.

In the following we combine the work of Love \cite{4} with that of Green and Zerna \cite{10}. We show that for two parallel coaxial disks of different radii the expansion in terms of complex potentials leads directly to a pair of coupled integral equations for the weight functions of the two disks. Hence the charge density and capacitance can be evaluated by a numerical procedure. For equal radii the Love integral equation follows as a special case.

\section{\label{II}Single disk}

We consider a circular disk of radius $a$ with axially symmetric charge distribution in infinite space. We use cylindrical coordinates $(r,\varphi,z)$ with the disk in the plane $z=0$ and centered at the origin. Laplace's equation for the electrostatic potential $\phi(\vc{r})$ reads
\begin{equation}
\label{2.1}\nabla^2\phi=-4\pi\rho=-4\pi\sigma(r)\delta(z),
\end{equation}
where $\rho(\vc{r})$ is the charge density, and $\sigma(r)$ is the surface charge density of the disk. The surface charge density determines the jump condition
\begin{equation}
\label{2.2}\frac{\partial\phi}{\partial z}\bigg|_{z=0+}-\frac{\partial\phi}{\partial z}\bigg|_{z=0-}=-4\pi\sigma(r).
\end{equation}

In order to solve Laplace's equation we use an expansion in terms of complex potentials \cite{4},\cite{10}. It is easily checked that the fundamental solution
\begin{equation}
\label{2.3}W(r,z,s)=\mathrm{Re}\frac{1}{\sqrt{r^2+(z-is)^2}}
\end{equation}
satisfies Laplace's equation everywhere, except on a disk of radius $s$ centered at the origin in the $z=0$ plane. The potential is continuous everywhere, and $W(r,0,s)=0$ for $0\leq r<s$. Its derivative with respect to $z$ has a discontinuity for $z=0$ and $0\leq r\leq s$. The relation to the solution by the alternative expansion in Bessel functions is provided by the identity\cite{11}
\begin{equation}
\label{2.4}W(r,z,s)=\int^\infty_0\cos ks\;J_0(kr)e^{-k|z|}\;dk.
\end{equation}
We do not need this in the following.

We solve the problem for charge density $\sigma(r)$ by linear superposition of fundamental solutions
 \begin{equation}
\label{2.5}\phi(r,z,a)=\int^a_0W(r,z,s)g(s)\;ds,
\end{equation}
with weight function $g(s)$. The value of the potential at the disk is \cite{5},\cite{10}
\begin{equation}
\label{2.6}\phi(r,0,a)=\int^r_0\frac{g(s)}{\sqrt{r^2-s^2}}\;ds,\qquad 0\leq r<a,
\end{equation}
since only  values $g(s)$ for $s\leq r$ contribute. We abbreviate $f(r)=\phi(r,0,a)$, and write the relation as
\begin{equation}
\label{2.7}f=\mathcal{A}g,\qquad 0\leq r<a,
\end{equation}
with linear operator $\mathcal{A}$ defined by the Abel type transform \cite{5}
\begin{equation}
\label{2.8}f(r)=\int^r_0\frac{g(s)}{\sqrt{r^2-s^2}}\;ds.
\end{equation}
The inverse transform $\mathcal{A}^{-1}$ is given by
\begin{equation}
\label{2.9}g(r)=\frac{2}{\pi}\frac{d}{dr}\int^r_0\frac{sf(s)}{\sqrt{r^2-s^2}}\;ds.
\end{equation}
It follows from Eqs. (2.2) and (2.5) that the surface charge density and the weight function are related by \cite{5},\cite{10}
\begin{equation}
\label{2.10}\sigma(r)=\frac{-1}{2\pi r}\frac{d}{dr}\int^a_r\frac{sg(s)}{\sqrt{s^2-r^2}}\;ds.
\end{equation}
The total charge is
\begin{equation}
\label{2.11}Q=2\pi\int^a_0r\sigma(r)\;dr=\int^a_0g(s)\;ds.
\end{equation}

We note that the Abel transform has the simple properties
\begin{equation}
\label{2.12}\mathcal{A}\;1=\frac{\pi}{2},\qquad\mathcal{A}^{-1}1=\frac{2}{\pi}.
\end{equation}
For a perfectly conducting disk the potential on the disk is a constant $V$, so that Eq. (2.7) becomes $V=\mathcal{A}g$. Applying the inverse operator $\mathcal{A}^{-1}$ we obtain in this case
\begin{equation}
\label{2.13}g(s)=\frac{2}{\pi}V,\qquad 0\leq s<a.
\end{equation}
From Eq. (2.10) one finds for the corresponding charge density
\begin{equation}
\label{2.14}\sigma(r)=\frac{V}{\pi^2}\frac{1}{\sqrt{a^2-r^2}}.
\end{equation}
The total charge is
\begin{equation}
\label{2.15}Q=CV,\qquad C=\frac{2}{\pi}\;a,
\end{equation}
where $C$ is the capacitance. These results are identical to those found for a thin oblate spheroid \cite{12}. The expression given by Landau and Lifshitz must be counted twice, since in their geometry both sides of the plate contribute equally.

The weight function $g(s)$, given by Eq. (2.13), yields for the potential
\begin{equation}
\label{2.16}\phi(r,z,a)=\frac{2V}{\pi}\mathrm{Re}\;i\log\frac{z-ia+\sqrt{r^2+(z-ia)^2}}{a}.
\end{equation}
The expression is more attractive than alternative ones in the form of a Bessel function integral \cite{5} or in terms of spheroidal coordinates \cite{12A}.

\section{\label{III}Circular plate condenser}

Next we consider two parallel circular plates of radii $a_1$ and $a_2$, both perpendicular to the $z$ axis and with center on the axis. We take the center of plate 1 to be at $(0,0,h)$ with height $h$, and the center of the second plate at the origin. The charge densities $\sigma_1(r)$ and $\sigma_2(r)$ are assumed to be radially symmetric. Hence the potential can be expressed as
\begin{equation}
\label{3.1}\phi(r,z)=\int^{a_1}_0W(r,z-h,s)g_1(s)\;ds+\int^{a_2}_0W(r,z,s)g_2(s)\;ds,
\end{equation}
with weight functions $g_1(s)$ and $g_2(s)$ related to the charge densities by Eq. (2.10).

If the plates are perfectly conducting the charge densities adjust such that the potential on each plate is a constant. Using Eq. (3.1) on each plate we find
\begin{eqnarray}
\label{3.2}\int^r_0\frac{g_1(s)}{\sqrt{r^2-s^2}}\;ds+\int^{a_2}_0W(r,h,s)g_2(s)\;ds&=&V_1,\nonumber\\
\int^{a_1}_0W(r,-h,s)g_1(s)\;ds+\int^r_0\frac{g_2(s)}{\sqrt{r^2-s^2}}\;ds&=&V_2.
\end{eqnarray}
We apply the inverse Abel transform $\mathcal{A}^{-1}$ to both equations. We note that
\begin{equation}
\label{3.3}\int^r_0\frac{sW(s,h,b)}{\sqrt{r^2-s^2}}\;ds=\mathrm{Re}\arctan\frac{r}{h-ib}.
\end{equation}
Using this we find the pair of equations
\begin{eqnarray}
\label{3.4}g_1(r)+\int^{a_2}_0K(r,s)g_2(s)\;ds&=&\frac{2}{\pi}V_1,\nonumber\\
\int^{a_1}_0K(r,s)g_1(s)\;ds+g_2(r)&=&\frac{2}{\pi}V_2,
\end{eqnarray}
with integral kernel
\begin{equation}
\label{3.5}K(r,s)=\frac{h}{\pi}\bigg[\frac{1}{h^2+(r-s)^2}+\frac{1}{h^2+(r+s)^2}\bigg].
\end{equation}

If the two disks have equal radii $a_1=a_2=a$, then the two weight functions $g_1(s)$ and $g_2(s)$ are related by symmetry. If the two potentials are equal $V_1=V_2=V$, then the two weight functions are also equal. If the potentials have opposite sign $V_1=-V_2=V$, then the two weight functions have also opposite sign, $g_1(s)=-g_2(s)=g(s)$. In that case the weight function $g(s)$ satisfies the Love equation
\begin{equation}
\label{3.6}g(r)-\int^a_0K(r,s)g(s)\;ds=\frac{2}{\pi}\;V.
\end{equation}
In the symmetric case the second term has the opposite sign.

In the antisymmetric case $V_1=-V_2=V$ the potential vanishes by symmetry at the midplane $z=h/2$. Hence the potential in the space $z>h/2$ is identical with that for the situation of a plate of radius $a$ at height $h$ and potential $V$ and an infinite conducting plate at height $h/2$ and vanishing potential. It follows from the second equation (3.4) that the weight function on the infinite plate is given by
\begin{equation}
\label{3.7}g_0(r)=-\int^a_0K_{1/2}(r,s)g(s)\;ds,
\end{equation}
where $K_{1/2}(r,s)$ is given by Eq. (3.5) with $h$ replaced by $h/2$, and $g(s)$ is the solution of Eq. (3.6).

\section{\label{IV}Discussion}

We have shown that the axially symmetric electrostatic problem of two parallel charged disks can be formulated efficiently by an expansion in complex potentials. Recourse to an expansion in terms of Bessel functions is not required. The first method leads directly to coupled integral equations for the weight functions on the two disks.

For the case where the radii are equal and the charge densities are equal and opposite in sign, the two coupled integral equations reduce to Love's integral equation. The integral equations can be solved numerically by various methods. Especially the case of a circular condenser with small separation of the plates has received much attention \cite{13},\cite{14}. Approximations for a large gap between the plates have also been studied \cite{15}. For intermediate values of the separation distance an expansion of the weight function in terms of a suitably chosen set of orthonormal functions can be used \cite{6},\cite{16}.

The above straightforward method of derivation of the pair of integral equations is of more general interest. Expansion in terms of Bessel functions and corresponding dual integral equations have been employed in several axially symmetric problems of low Reynolds number hydrodynamics \cite{17}-\cite{20}. It will be worthwhile to search for a simplified derivation of the resulting integral equations for the weight functions by direct expansion in complex fundamental solutions.

\end{document}